\documentclass[journal]{IEEEtran}
\usepackage{amsmath}
\usepackage{graphicx}

\ifCLASSINFOpdf
\else
\fi

\newcommand{\xiv}{\mbox{\boldmath$\xi$}}

\begin{document}

\title{CPIC: a Curvilinear Particle-In-Cell code for
plasma-material interaction studies}

\author{Gian~Luca~Delzanno,
        Enrico~Camporeale,
J.~David~Moulton,
Joseph~E.~Borovsky,
Elizabeth~A.~MacDonald
        and~Michelle~F.~Thomsen
\thanks{G.L. Delzanno, E. Camporeale, and J.D. Moulton are with the T-5 Applied Mathematics and Plasma Physics,
Los Alamos National Laboratory, Los Alamos,
NM, 87545 USA e-mail: delzanno@lanl.gov, enrico@lanl.gov, moulton@lanl.gov.}
\thanks{J.E. Borovsky is with the Space Science Institute, Boulder, CO, 80301 USA e-mail: 	
jborovsky@SpaceScience.org.}
\thanks{E.A. MacDonald and M.F. Thomsen are with the ISR-1 Space Science Applications,
Los Alamos National Laboratory, Los Alamos,
NM, 87545 USA e-mail: macdonald@lanl.gov, mthomsen@lanl.gov.}
\thanks{Manuscript received ????, ????; revised ????, ????.}}

\markboth{IEEE TRANSACTIONS ON PLASMA SCIENCE,~Vol.~?, No.~?, ????}%
{Shell \MakeLowercase{\textit{et al.}}: Bare Demo of IEEEtran.cls for Journals}

\maketitle

\begin{abstract}
We describe a new electrostatic Particle-In-Cell (PIC) code in curvilinear geometry called Curvilinear PIC (CPIC). 
The code models the microscopic (kinetic) evolution of a plasma with the PIC method, coupled with an adaptive computational 
grid that can conform to arbitrarily shaped domains. CPIC is particularly suited for multiscale problems associated with 
the interaction of complex objects with plasmas. A map is introduced between the physical space and the logical space, 
where the grid is uniform and Cartesian. In CPIC, most operations are performed in logical space. CPIC was designed following 
criteria of versatility, robustness and performance. Its main features are the use of structured meshes, a scalable field solver based on the black box 
multigrid algorithm and a hybrid mover, where particles' position is in logical space while the velocity is in physical space. 
Test examples involving the interaction of a plasma with material boundaries are presented.
\end{abstract}


\IEEEpeerreviewmaketitle

\section{Introduction}

\IEEEPARstart{P}{article}-In-Cell (PIC) codes have been the main tool for plasma physics kinetic 
simulations for several decades \cite{birdsall,hockney}. In the PIC algorithm, a number of macroparticles 
(each representing many particles of the physical system) move through a computational grid due 
to the electromagnetic fields. The latter are self-consistently calculated from the particles on the grid. 
It is the interplay between the particles and the grid that makes PIC an efficient algorithm: the calculation 
of the force on the particles scales linearly with the number of particles (assuming that these are many more 
than the number of grid cells), as opposed to the quadratic scaling typical of grid-free molecular dynamics methods, 
where the particle force is calculated summing the interaction with every other particle \cite{hockney}.

Traditionally, PIC codes use a uniform, Cartesian grid and integrate the particle orbit by using explicit schemes. 
As such, they need to resolve the shortest length scale and the fastest frequency of the plasma for numerical stability reasons. 
Since plasmas are inherently multiscale, it follows that PIC methods require a significant amount of computational resources to 
handle this disparity of scales. (We note that PIC methods with implicit time stepping, such as
the implicit moment method PIC \cite{brackbill&forslund} or the fully implicit PIC methods recently developed in Refs. \cite{chacon,markidis},
avoid the stability constraints typical of explicit PIC.)

The problems of PIC for multiscale plasma simulations can become worse if one needs to simulate the interaction of a plasma with 
material objects (for instance spacecraft, dust grains or the walls of laboratory devices), since the objects introduce additional 
scales in the system. As an example, consider a spacecraft at geosynchronous orbit (an altitude of approximately 36,000 km above the Earth's surface). 
The typical length scales are the spacecraft characteristic size $L_{spacecraft}\sim 1-10$ m, 
the plasma Debye length $\lambda_{D}\sim 250$ m, 
the electron gyroradius $\rho_e\sim 750$ m and the proton gyroradius $\rho_i\sim 30$ km. One quickly realizes that attempting to simulate such systems 
with a uniform, Cartesian grid explicit PIC in three dimensions, and with grid size set by the spacecraft, is unfeasible even on today's supercomputers.
We note that some approaches \cite{roussel98,hutchinson2}
assume Boltzmann electrons and therefore only need to resolve the ions scales. These approaches do not have the cell size constraints of the explicit, full PIC method, and are 
typically applicable when the material object is negatively charged and in absence of potential barriers.

Focusing on spatial scales, this discussion points to the importance of introducing some kind of grid adaptivity in the  PIC method, 
in order to handle efficiently (1) complex geometries and/or (2) the interaction of plasmas with objects whose characteristic size is much 
smaller than typical plasma length scales. Traditionally, PIC with non-uniform/adaptive meshes has followed two paths. One is the use 
of adaptive mesh refinement (AMR) techniques \cite{vay}. The other is the use of body-fitted grids, namely grids that conform exactly to a 
given surface \cite{westermann1} and avoid the stair-stepping typical of AMR grids.

In this paper, we focus on the body-fitted PIC approach. References \cite{westermann1,westermann2} and \cite{munz} used this approach to model pulsed-power diode devices with 
complex geometries. In Refs. \cite{westermann1,westermann2} and \cite{munz}, a coordinate transformation (i.e. a map) from the physical space to the logical space, where the 
grid is uniform and Cartesian, is introduced. The quasi-stationary Maxwell equations are solved with finite differences in logical space, 
while particle orbits are integrated in physical space. Since the grid is distorted in physical space, a tracking algorithm to locate in 
which cell a particle resides must be used \cite{westermann3}. Eastwood et al. \cite{eastwood} follow the body-fitted PIC approach for modeling microwave devices. They use a finite 
element formulation and solve both Maxwell's equations and the particle orbits in logical space. (See also the recent work by Fichtl et al. 
\cite{fichtl} on an electrostatic PIC code completely designed in logical space with a particle orbit integrator that preserves phase-space area.) 
Wang et al. \cite{wang}, on the other hand, update Maxwell's equations in 
physical space but use a hybrid orbit integrator (where the particle position is in logical space and the velocity is in physical space). 
We also note that in the field of spacecraft-plasma interaction there are several efforts currently being developed that can be classified as 
body-fitted PIC. These are NASCAP-2k \cite{mandell}, SPIS \cite{roussel} and PTetra \cite{marchand}. While the specific details of the numerical implementations differ 
(see Refs. \cite{mandell,roussel,marchand}), in general these approaches are in the electrostatic limit and Poisson's equation is solved with an iterative solver 
(conjugate gradient or GMRES method) with some form of preconditioning. Furthermore, they do not introduce the logical space and all the operations 
are performed in physical space. Some of these approaches (SPIS and PTetra) use unstructured computational meshes.
The other major code, MUSCAT \cite{muranaka}, is not body-fitted: it uses a structured, Cartesian, uniform mesh and Poisson's 
equation is solved with the Fast Fourier Transform (FFT) algorithm, using parallel domain decomposition on multiple processors.

In this paper, we present a fully-kinetic, electrostatic, body-fitted PIC code in general curvilinear geometry called Curvilinear PIC (CPIC). 
While the general formulation of the method follows earlier works \cite{westermann1,eastwood,wang}, CPIC was designed with flexibility, robustness and performance 
considerations as the main target. For this reason, its main features are
\begin{enumerate}
\item The use of structured computational meshes. Modern multi- and many-core computer architectures achieve their best
performance when data movement is minimized, and computations can be expressed in a data parallel fashion without indirect memory access
patterns.  These characteristics give a significant advantage to structured grid approaches that are able to take advantage of known
direct access patterns, simpler geometric relationships, and generally lower computational complexity.
For instance, the Black Box Multigrid (BoxMG) algorithm used in CPIC takes advantage of both known data access patterns and the bounded complexity of coarse-grid 
operators to solve general Poisson problems approximately 10 times faster than algebraic multigrid methods (AMG) \cite{maclachlan}.
\item The field solver. We use the BoxMG algorithm \cite{dendy,dendy2,moulton} as solver. 
This algorithm is scalable (namely optimal): the computational cost to converge to the problem solution is 
linearly proportional to the number of grid cells. This means that, at least theoretically, there is no loss of performance when the problem 
size increases, as typical of unpreconditioned iterative solvers.
In addition, robust variationally-based structured grid methods such as BoxMG \cite{dendy,dendy2} are able to solve problems with strongly discontinuous coefficients on the 
logically structured (i.e., body fitted) grids needed here.  
\item The particle mover. We use the hybrid mover of Ref. \cite{wang}. We have compared the performance of the hybrid mover to the (more common) physical 
space mover on test cases and found that the hybrid mover is typically faster and much more robust. Indeed, it avoids the issue of particle tracking 
needed by the physical space mover, and it also deals more efficiently with complex geometries, without the need of an extra routine to assess whether 
a particle crosses the domain boundaries. 

\end{enumerate}

The paper is organized as follows. In Section II, we describe the main algorithmic and implementation aspects of CPIC. 
In Section III, we present some benchmark tests involving the interaction of a plasma with material boundaries. In Section IV, we draw conclusions.

\section{Overview of the curvilinear PIC (CPIC)}

We model a collisionless, magnetized plasma described by Vlasov's equations
\begin{equation}
 \frac{\partial f_\alpha}{\partial t}+{\bf v} \cdot \nabla f_\alpha+\frac{q_\alpha}{m_\alpha}\left({\bf E}+{\bf v} \times {\bf B}_0 \right)\cdot
\nabla_{v} f_\alpha=0
\label{f}
\end{equation}
where the subscript $\alpha$ refers to the plasma species, $\alpha=e,\,(i)$ for electrons (ions). 
In Eq. (\ref{f}), $f_\alpha({\bf x}, {\bf v}, t)$ is the plasma distribution function,
${\bf E}=-\nabla \phi$ is the electric field  ($\phi$ is the electrostatic potential),
${\bf B}_0({\bf x})$ is the  magnetic field, while $q_\alpha$ and $m_\alpha$ are the charge and the mass of the plasma particles. 
We focus on the electrostatic limit, therefore Eq. (\ref{f}) is coupled through the electrostatic potential via Poisson's equation
\begin{equation}
 \nabla^2 \phi =-\frac{\sum q_\alpha n_\alpha}{\varepsilon_0}=-\frac{\rho}{\varepsilon_0},
\label{pois}
\end{equation}
where the plasma densities are given by
\begin{equation}
 n_\alpha=\int f_\alpha d {\bf v},
\label{dens}
\end{equation}
and the magnetic field ${\bf B}_0$ does not evolve in time. In Eq. (\ref{pois}), $\varepsilon_0$ is the permittivity of vacuum.

The PIC method solves Eq. (\ref{f}) numerically by introducing macroparticles, each representing a large number of actual particles of the system, 
and following their characteristics:
\begin{eqnarray}
 && \frac{d {\bf x}_p}{dt}={\bf v}_p, \label{xp} \\
&& m_p \frac{d {\bf v}_p}{dt}=q_p\left[{\bf E}\left({\bf x}_p \right)+{\bf v}_p \times {\bf B}_0\left({\bf x}_p \right) \right], \label{vp}
\end{eqnarray}
where ${\bf x}_p$ and ${\bf v}_p$ are position and velocity of each particle. 
Hence, a standard electrostatic PIC cycle, repeated at each time step of the simulation, consists of the following four steps:
\begin{enumerate}
 \item \underline{Particle mover}: particles are moved according to Eqs. (4) and (5), given the electromagnetic fields.
\item \underline{Particle-to-Grid (P$\rightarrow$G) interpolation}: the charge carried by the particles is accumulated on the computational grid via interpolation, 
to obtain the plasma charge density.
\item \underline{Field solver}: Poisson's equation is solved on the computational grid, given the charge density.
\item \underline{Grid-to-Particle (G$\rightarrow$P) interpolation}: the electromagnetic fields are interpolated to the particle position.
\end{enumerate}
We now proceed to discuss each of these steps as they are treated in CPIC.

\subsection{Curvilinear formulation}\label{curv}
First, we introduce a coordinate transformation
\begin{equation}
 {\bf x}={\bf x}\left(\xiv \right)
\label{map}
\end{equation}
between the physical space $X$ with coordinates ${\bf x}=\left(x^1,\,x^2,\,x^3 \right)$, and the logical space $\Xi$ (the unit cube), 
with coordinates $\xiv=\left(\xi^1,\,\xi^2,\,\xi^3 \right)=\left(\xi,\,\eta,\,\zeta \right)$.
Given a computational mesh in physical space corresponding to the simulation domain, Eq. (\ref{map}) maps it to the unit cube in logical space, 
where the computational mesh is uniform and Cartesian.
We define the Jacobi matrix as \cite{liseikin}
\begin{equation}
 j_{\alpha\beta} \left(\xiv\right)=\frac{\partial x^\alpha}{\partial \xi^\beta},\,\,\,\alpha,\,\beta=1,\,2,\,3,
\label{jcov}
\end{equation}
while its inverse is
\begin{equation}
 k^{\alpha\beta} \left({\bf x}\right)=\frac{\partial \xi^\alpha}{\partial x^\beta},\,\,\,\alpha,\,\beta=1,\,2,\,3.
\label{jcon}
\end{equation}
The Jacobian of the transformation is the determinant of the Jacobi matrix,
\begin{equation}
 J\left(\xiv\right)={\rm det}\left[j_{\alpha\beta} \right].
\label{jac}
\end{equation}
Similarly, the covariant metric tensor is defined as \cite{liseikin}
\begin{equation}
 g_{\alpha\beta}\left(\xiv\right)=\frac{\partial x^\gamma}{\partial \xi^\alpha}\frac{\partial x^\gamma}{\partial \xi^\beta},\,\,\,\alpha,\,\beta=1,\,2,\,3,
\label{gcov}
\end{equation}
while the contravariant metric tensor is
\begin{equation}
 g^{\alpha\beta}\left({\bf x}\right)=\frac{\partial \xi^\alpha}{\partial x^\gamma}\frac{\partial \xi^\beta}{\partial x^\gamma},\,\,\,\alpha,\,\beta=1,\,2,\,3.
\label{gcon}
\end{equation}
Here and everywhere else in the text, repeated indices imply summation (Einstein's notation). 
Note that the metric tensor matrix is symmetric and that grids are orthogonal if $g_{12}=g_{13}=g_{23}=0$.
With these quantities, we can express all the metric elements from physical space to logical space.
For instance, Poisson's equation in logical space becomes
\begin{equation}
 \nabla^2 \phi=\frac{1}{\Omega J} \frac{\partial}{\partial \xi^i}\left(\Omega J g^{ij} \frac{\partial \phi}{\partial \xi^j} \right)=-\frac{\rho}{\varepsilon_0},
\label{poiscurv}
\end{equation}
where we have introduced a geometric factor $\Omega$ that can be used when the geometry of the physical space is non-Cartesian. 
For instance, for Cartesian geometry we have $\left(x^1=x,\,x^2=y,\,x^3=z\right)$ and $\Omega=1$, while for cylindrical geometry we have
$\left(x^1=r,\,x^2=\theta,\,x^3=z\right)$ and $\Omega=r$.

From the discussion so far, it is clear that the construction of the map from physical to logical space, Eq. (\ref{map}), is a critical step in CPIC.
In the case of extremely simple geometries, like the test examples considered in this paper, one can specify such map analytically. Alternatively,
one can use a suitable mesh generator. For cases that are still relatively simple, where the map is characterized by a single or few structured blocks,
we normally use methods based on the solution of a set of partial differential equations such as Winslow's method \cite{winslow-unpublished-var-diff} 
or the optimal distortion method \cite{delzanno08}. On the other hand, in order to capture the full complexity of a spacecraft we plan to
use commercially available software for structured meshes, typically developed in the computational fluid dynamics community (see for instance GridPro, https:/gridpro.com).
The resulting computational grid is block-structured and, for complex geometries, might involve hundreds of blocks. This will require upgrading CPIC to handle block-structured meshes.
However, the fact that each block maps to a logically rectangular
Cartesian mesh ensures that the localization of the particles on the mesh of each block and the particle mover remain essentially unchanged, while there is an additional computational
cost in assigning each particle to a block. In CPIC this last step is trivial provided that each grid
block stores the maximum and minimum value of its logical coordinates (as done in the domain decomposition for the parallelization),
since the particle position is in logical space. We also note that in CPIC the metric coefficients (the contravariant metric tensor, the inverse
of the Jacobi matrix and the Jacobian) are computed from finite difference approximation given the discrete map (\ref{map}) known at cell vertices (see for instance \cite{delzanno08}).
There is no conceptual difference or additional computational cost between a structured or block-structured mesh since the metric coefficients are defined in each block.
Furthermore, there are three popular approaches to developing a Poisson solver for block-structured grids.  
First, one may use a domain decomposition approach, with BoxMG as the sub-domain solver.  This shares convergence properties of domain decomposition methods, and for large problems would require overlap to obtain satisfactory scaling.  
Second, BoxMG could be extended following the approach used in the HYPRE libraries Semi-Structured Grid interface to its parallel Semi-Coarsening Multigrid solver, which manages the coarsening of each block independently, while maintaining the full connectivity of the global problem \cite{hypre}.
Finally, a bounding box approach can be used to include the entire global domain in a single BoxMG solve, with internal boundary conditions used at the block interfaces, and this is the approach that we plan to adopt.

\subsection{Computational aspects of CPIC}

In CPIC quantities are expressed in normalized units: lengths are normalized to the electron Debye length, 
velocities to the electron thermal velocity, 
time to the electron plasma frequency, the electrostatic potential to the electron temperature, densities to a reference density and the 
magnetic field to a reference magnetic field. In the tests presented throughout the paper, we will always use normalized quantities.

CPIC is designed in a staggered formulation, where the electrostatic potential and the density are at cell centers (labeled with subscript c) 
and the electric field is at vertices (v). The staggered formulation and the fact that the assignment function for the interpolation 
from particles to the grid is one order higher than the one for the interpolation from the grid to the particles (see below) imply that CPIC is 
an energy-conserving PIC as defined in Ref. \cite{birdsall}.

We have developed and successfully tested a two-dimensional (2D) and three-dimensional (3D) version of CPIC. CPIC2D is fully parallelized with
domain decomposition and the MPI library to handle the communication among processors. Parallelization of CPIC3D is ongoing.

\subsubsection{P$\rightarrow$G interpolation}
The accumulation of the particle charge on the grid is given by
\begin{equation}
 n_{\alpha,\,c}=\frac{\sum_p q_p W_{pc} \left(\xiv_p-\xiv_c \right)}{J_c \Delta \xi \Delta \eta \Delta \zeta}
\label{pg}
\end{equation}
where the assignment function is
\begin{equation}
 W_{pc}=b_2\left(\xi_p-\xi_c\right)b_2\left(\eta_p-\eta_c\right)b_2\left(\zeta_p-\zeta_c\right)
\label{Wpc}
\end{equation}
and $b_2$ is the b-spline of order 2:
\begin{equation}
 b_2\left(\xi \right)=\left\lbrace
  \begin{aligned}
   &\frac{3}{4}-\xi^2, &|\xi|<\frac{1}{2},\\
   &\left(\frac{3}{2}-|\xi|\right)^2, &\frac{1}{2}<|\xi|<\frac{3}{2},\\
   & 0, & {\rm otherwise}.
  \end{aligned}
\right.
\label{b2}
\end{equation}
In Eq. (\ref{pg}), $\sum_p$ implies summation over all the particles.

\subsubsection{G$\rightarrow$P interpolation}
The interpolation of the force field from the computational mesh to the particle position is given by
\begin{equation}
{\bf E}_p=\sum_v {\bf E}_v W_{vp} \left(\xiv_v-\xiv_p \right)
\label{vvp}
\end{equation}
with the assignment function
\begin{equation}
 W_{vp}=b_1\left(\xi_v-\xi_p\right)b_1\left(\eta_v-\eta_p\right)b_1\left(\zeta_v-\zeta_p\right)
\label{Wvp}
\end{equation}
given as the product of b-splines of order 1
\begin{equation}
 b_1\left(\xi \right)=\left\lbrace
  \begin{aligned}
   &1-|\xi|, &|\xi|<1,\\   
   & 0, & {\rm otherwise}.
  \end{aligned}
\right.
\label{b1}
\end{equation}
We note that the interpolation operations are performed in logical space and therefore are equivalent to what is done in the standard PIC with uniform grid.
The same approach is used in Ref. \cite{lapenta}.

\subsubsection{Field solver} 
We solve Poisson's equation on the logical grid. We use a second-order, conservative discretization and impose 
boundary conditions via ghost cells. The resulting linear system is solved with a robust variational multigrid method \cite{dendy}. 
The need for a scalable algorithm motivates the use of a multigrid solver, as it is one of very few truly scalable solvers 
(it can be shown that the order of complexity of the method is $O(N)$, with $N$ the number of unknowns of the linear system \cite{Briggs}). Moreover, 
efficient parallel implementations are readily available.

Multigrid methods are iterative, and achieve their efficiency through the recursive use of successively coarser discrete problems 
(i.e., a sequence of coarse-grid discrete operators) in conjunction with smoothing on each level (e.g., a single Gauss-Seidel iteration on each level) 
to damp the highly oscillatory errors associated with each grid \cite{Briggs}. Early work on multigrid began with structured grids and took a natural but 
simple geometric viewpoint: coarsening removed every other point in each coordinate direction; the same discretization was applied directly at 
each grid resolution; and the interpolation and restriction operators were defined based on the grid geometry alone. Unfortunately, this approach was not 
robust for challenging problems with discontinuous coefficients or severely distorted grids. To address this shortcoming, a robust approach for problems 
on logically structured grids was developed that only requires the user to provide the fine-grid matrix and the right-hand side \cite{dendy}. This Black Box Multigrid 
(BoxMG) algorithm uses a variational coarse-grid operator, as it minimizes the error in the range of the interpolation. Moreover, it uses the entries in 
the matrix (the discrete operator) to define the interpolation, dubbed operator-induced interpolation. This approach ensures that important properties of 
the fine-scale PDE are well approximated at all resolutions, and leads to a robust scalable solver suitable for grids of any dimension, discontinuous 
coefficients, and any type of boundary condition (Dirichlet, Neumann, Robin, and periodic) \cite{moulton}. In addition, it is robust for anisotropic problems, 
provided line-relaxation is used instead of point smoothing.

As an illustration of this capability, we use the method of manufactured solutions \cite{roache} on the (normalized) Poisson equation in spherical 
geometry in two-dimensions (2D). This can be done by assuming that the physical space corresponds to the cylindrical geometry
$\left(x^1=r,\,x^2=\theta,\,x^3=z\right)$, with $\partial/\partial \theta=0$, and introducing the following coordinate transformation
\begin{eqnarray}
 && r=\left[r_1+\left(r_2-r_1 \right)\xi \right]\sin \left[\pi\left(1-\eta\right) \right], \label{r} \\
&& z=\left[r_1+\left(r_2-r_1 \right)\xi \right]\cos \left[\pi\left(1-\eta\right) \right], \label{z}
\end{eqnarray}
where $r_1$ and $r_2$ are the inner and outer radii of the physical domain (we choose $r_1=1$ and $r_2=10$ in this example). 
The geometric factor is set to $\Omega=r$. Next, we seek the following solution
\begin{equation}
 \phi\left(\xi,\,\eta \right)=\sin\left[\left(r_2-r_1\right)^2 \xi \left(1-\xi\right) \right]\cos \left(\pi \eta^2 \right),
\label{mms}
\end{equation}
which is inserted in Poisson's equation to obtain the density. We then solve Poisson's equation with the BoxMG solver on a serial machine. 
We use line-relaxation since this example is fairly anisotropic in terms of metric tensor coefficients. The results are presented in Table I, 
in the form of a convergence study changing the number of grid points. The second column of Table \ref{t1} shows the $L_2$ norm of the error of the 
numerical solution relative to the analytical solution in Eq. (\ref{mms}), while the third column is the inverse of the ratio of the error 
calculated on a given grid divided by the error from the previous coarser grid. The second-order convergence of our discretization scheme is 
clearly recovered (i.e. when doubling the resolution, the error ratio is equal to 4 for sufficiently large grids). The fourth column shows 
the number of iterations needed by the solver to converge, with relative tolerance set to $r_{tol}=10^{-10}$. The iteration count remains practically constant. 
The fifth and sixth columns show the time to reach convergence and the ratio of such time on a given grid relative to the previous coarser grid. 
Here we have only plotted the time for the more refined grids in order to identify the asymptotic scaling. As expected, it shows that the 
algorithm is scalable: doubling the grid resolution in each direction requires about four times more time for convergence. 
This is also consistent with the iteration count remaining constant as the grid is refined.

\begin{table}[!t]
\caption{Convergence study of the BoxMG solver changing the number of grid points.}
\label{t1}
\centering
\begin{tabular}{|c|c|c|c|c|c|}
\hline
Grid & $L_2$ error & Error & Number of & Time [s] & Time \\ 
 &  & ratio & iterations & & ratio \\ \hline \hline
$16^2$ & $5.3 \cdot 10^{0}$ & & $5$ & & \\ \hline
$32^2$ & $2.0 \cdot 10^{-1}$ & $28.6$ & $5$ & & \\ \hline
$64^2$ & $3.7 \cdot 10^{-2}$ & $5.1$ & $5$ & & \\ \hline
$128^2$ & $8.8 \cdot 10^{-3}$ & $4.1$ & $5$ & $0.06$ & \\ \hline
$256^2$ & $2.2 \cdot 10^{-3}$ & $4.0$ & $6$ & $0.27$ & $4.5$ \\ \hline
$512^2$ & $5.4 \cdot 10^{-4}$ & $4.0$ & $6$ & $1.2$ & $4.4$ \\ \hline
$1024^2$ & $1.3 \cdot 10^{-4}$ & $4.0$ & $5$ & $4.2$ & $3.4$ \\ \hline
$2048^2$ & $3.4 \cdot 10^{-5}$ & $4.0$ & $5$ & $16.6$ & $4.0$ \\ \hline
$4096^2$ & $8.4 \cdot 10^{-6}$ & $4.0$ & $5$ & $66.4$ & $4.0$ \\ \hline \hline
\end{tabular}
\end{table}

\subsubsection{Particle mover}
We initially specialize the discussion to the case of Cartesian physical space. The most common approach for particle mover 
in PIC codes with non-orthogonal grids is to move particles in physical space via Eqs. (\ref{xp}) and (\ref{vp}). As the particles move through a 
distorted mesh, a tracking/localization procedure is  required in order to locate the particle in a cell, as needed by the interpolation and 
accumulation routines. (An alternative, used in Ref. \cite{lapenta}, is to move particles in physical space and then invert the map
${\bf x}\left(\xiv\right)$ to obtain the particles' position in logical space. This technique can be very efficient for cases where the map can 
be inverted analytically.) In addition, the time discretization of Eqs. (\ref{xp}) and (\ref{vp}) is usually explicit, with the leap-frog integrator 
\cite{birdsall} as the most popular choice. The leap-frog algorithm staggers particles position and velocity by half time step, and is second-order accurate in time.
In CPIC, however, we follow a different approach which was proposed in Ref. \cite{wang}: particles retain their physical space velocity 
but are characterized by their logical space position (and therefore move in logical space). In other words, our mover consists of Eq. (\ref{vp}), 
while Eq. (\ref{xp}) is replaced by its logical space equivalent (obtained by projecting Eq. (\ref{xp}) on the contravariant base vector):
\begin{equation}
 \frac{d \xi_p^i}{dt}=k^{ij}v_p^j,\,\,\,i=1,\,2,\,3,
\label{hybridx}
\end{equation}
where the inverse of the Jacobi matrix is given by Eq. (\ref{jcon}) and we have used index notation. We refer to Eqs. (\ref{vp}) and 
(\ref{hybridx}) as the hybrid mover.

The hybrid mover offers some clear advantages over a physical space mover, mainly because of its simplicity:
\begin{enumerate}
\item There is no need of a tracking algorithm to locate the particles: a particle is immediately assigned to a cell through its logical space coordinate.
\item While tracking algorithms applied to complex boundaries (multiply connected or with concave boundaries) are typically not robust, this is not 
the case for the hybrid mover. In fact, a particle trivially leaves the domain if one of its logical coordinates becomes less than zero or greater than one.
\end{enumerate}
There is, however, an important disadvantage of the hybrid mover relative to its physical space counterpart: a simple leap-frog integrator 
loses second order accuracy on a non-orthogonal mesh. The simplest way to fix this problem and regain second order accuracy is to use a leap-frog integrator modified with a 
predictor-corrector approach \cite{wang}. The algorithm is as follows (index $n$ refers to the time level, $\Delta t$ is the time step):
\begin{enumerate}
 \item Perform a full-step update of the particle velocity to obtain ${\bf v}_p^{n+1/2}$:
\begin{equation}
 \frac{{\bf v}_p^{n+1/2}-{\bf v}_p^{n-1/2}}{\Delta t}=\frac{q_p}{m_p}\left[{\bf E}\left(\xiv_p^n \right)
+{\bf v}_p^n\times {\bf B}_0 \left(\xiv_p^n \right) \right],
\label{step1}
\end{equation}
with
\begin{equation}
 {\bf v}_p^n=\frac{{\bf v}_p^{n+1/2}+{\bf v}_p^{n-1/2}}{2}.
\label{step1a}
\end{equation}
This step is performed explicitly with the algorithm introduced by Boris \cite{boris}
which separates the velocity update due to the electric field from the update due to the magnetic field.
\item Predictor: perform a half-step update of the particle logical position to obtain $\xiv_p^\prime$:
\begin{equation}
 \frac{\xi_p^{i\prime}-\xi_p^{i,n}}{\displaystyle{\frac{\Delta t}{2}}}=k^{ij}\left(\xiv_p^n \right)v_p^{j,n+1/2},\,\,\,i=1,\,2,\,3.
\label{step2}
\end{equation}
\item Re-evaluate the metric coefficients at $\xiv_p^\prime$ via interpolation:
\begin{equation}
 k^{ij}\left(\xiv_p^\prime\right)=\sum_v k^{ij}\left(\xiv_v \right)W_{vp} \left(\xiv_v-\xiv_p^\prime \right).
\label{step3}
\end{equation}

\item Corrector: perform a full-step update of the particle logical position with the new metric coefficients, to obtain $\xiv_p^{n+1}$:
\begin{equation}
 \frac{\xi_p^{i,n+1}-\xi_p^{i,n}}{\Delta t}=k^{ij}\left(\xiv_p^\prime \right)v_p^{j,n+1/2},\,\,\,i=1,\,2,\,3.
\label{step4}
\end{equation}
\end{enumerate}

With these considerations, an important question is whether the hybrid mover is more or less efficient relative to the physical space mover 
from a performance standpoint. To answer this question, we have compared the hybrid mover with a physical space mover for the following test. 
We consider a square domain in 2D (Cartesian geometry), with the following physical-to-logical space map
\begin{eqnarray}
 &&x=\xi+\varepsilon \sin\left(2 \pi \xi\right)\sin\left(2 \pi \eta\right), \nonumber \\
&&y=\eta+\varepsilon \sin\left(2 \pi \xi\right)\sin\left(2 \pi \eta\right). \label{map2}
\end{eqnarray}
The parameter $\varepsilon$ controls the level of distortion of the grid. We fix the electromagnetic fields on the grid, the mesh size 
$\left(64 \times 64\right)$, the number of particles per cell $N_{p,\,cell}=100$ and we move the particles for $10^4$ time steps. 
The boundary conditions on the particles are periodic on all sides. The physical space mover requires a particle tracking and localization 
technique and we use the method proposed in Ref. \cite{westermann3}. The results of our comparison changing time step are presented in Tables \ref{t2} and \ref{t3}
for orthogonal (uniform) $\left(\varepsilon=0\right)$ and non-orthogonal $\left(\varepsilon=0.1\right)$ grids. 
For the latter, the ratio of the largest to smallest cell size is $4.4$. Tables \ref{t2} and \ref{t3} 
show the average time per particle per time step to complete the simulation, normalized to the time required by the hybrid mover on the 
uniform mesh with $\Delta t=0.001$. Let us focus first on Table \ref{t2}. 
We see that the time required by both movers to complete the simulation increases with the time step. 
For the hybrid mover, this is because additional operations are performed due to particles crossing the domain boundaries. For the physical space mover, 
however, there is extra work associated with the tracking/localization algorithm: with larger time steps, particles move further away and there are more particles 
crossing cell boundaries within the domain. Thus, for realistic time steps used in kinetic simulations, even on a uniform grid the hybrid mover can be faster. 
Let us now look at the results on the non-uniform grid, Table \ref{t3}. We notice that the average time required by the hybrid mover is comparable to the one on 
the uniform grid. This result is not surprising since, given the metric coefficients, the hybrid mover treats all the grids equally. It indicates, however, 
that the hybrid mover is extremely robust. The physical space mover, on the other hand, requires more time relative to its performance on the uniform grid. 
Again, this is not surprising as now the tracking/localization algorithm has to deal with distorted meshes. On this example, with time step $\Delta t=0.05$ the hybrid mover 
is about $60\%$ faster. 
We note that the results presented in Tables \ref{t2} and \ref{t3} involve the time spent by the algorithm to move and localize the particles
and the time spent to apply the (periodic) boundary conditions when a particle crosses the domain boundaries. The latter time is the same for the
hybrid and physical space movers since the example considered consists of a simple square domain. Therefore the performance loss of the physical space
mover is due to the particle tracking/localization only.
A further point worth noting is that particle tracking/localization algorithms are notoriously less robust on complex boundaries, 
and this will make the comparison even more favorable to the hybrid mover.

\begin{table}
\caption{Comparison of the time (per particle per time step) required by the hybrid mover and the physical space 
mover on a orthogonal (uniform) grid for various time steps.}
\label{t2}
\centering
\begin{tabular}{|c|c|c|}
\hline
Time step & Hybrid mover: & Physical space mover: \\
& time & time \\
\hline \hline
$0.001$ & $1.00$ & $0.80$  \\ \hline
$0.002$ & $1.02$ & $0.86$ \\ \hline
$0.004$ & $1.08$ & $1.05$ \\ \hline
$0.01$ & $1.17$ & $1.31$ \\ \hline
$0.025$ & $1.17$ & $1.31$ \\ \hline
$0.05$ & $1.17$ & $1.44$ \\ \hline \hline
\end{tabular}
\end{table}

\begin{table}
\caption{Comparison of the time (per particle per time step) required by the hybrid mover and the physical space 
mover on a non-orthogonal grid $\left(\varepsilon=0.1\right)$ for various time steps.}
\label{t3}
\centering
\begin{tabular}{|c|c|c|}
\hline
Time step & Hybrid mover: & Physical space mover: \\
& time & time \\
\hline
\hline
$0.001$ & $1.04$ & $1.04$  \\ \hline
$0.002$ & $1.06$ & $1.14$ \\ \hline
$0.004$ & $1.10$ & $1.31$ \\ \hline
$0.01$ & $1.16$ & $1.71$ \\ \hline
$0.025$ & $1.16$ & $1.71$ \\ \hline
$0.05$ & $1.17$ & $1.84$ \\ \hline \hline
\end{tabular}
\end{table}

One final comment is in order with regard to the application of CPIC to non-Cartesian geometry.
This is useful for instance when one wants to simulate the interaction of a plasma and an object
in a spherically symmetric system, and we will show some of these examples in Sec. \ref{tests}. 
In this case one can run CPIC in 2D cylindrical geometry
and impose azimuthal symmetry, thus allowing a much higher resolution than in a fully 3D simulation. 
From the perspective of the solver, this implies
setting $\Omega=r$ and $\partial/\partial \theta=0$ in Poisson's equation.
For the mover, on the other hand, one must take into account the contribution of the inertial
forces to the particle motion. This can be done with the cylindrical version of the Boris mover
\cite{boris}, suitably modified in its hybrid mover version. In addition, for simulations
involving a uniform background magnetic field, this algorithm can be improved to resolve the
particle gyromotion exactly \cite{patacchini,delzanno_mover}. Its importance resides in the fact
that, by describing the gyromotion exactly, one can relax the time step constraints that a
PIC code with a standard leap-frog mover would have when simulating a strongly magnetized plasma
with $\omega_{ce}\gg\omega_{pe}$ (with $\omega_{ce}$ the electron cyclotron frequency) and
deliver a significant computational savings.

\section{Tests}\label{tests}

We have successfully performed several tests to check the validity of CPIC against known solutions. 
These tests include Langmuir waves, Landau damping and two-stream instabilities, for cases where the plasma is not in contact with a material wall. 
Here, we present some additional tests where the plasma is bounded by a material boundary. These tests are performed in 2D (for the fields, while the particles 
retain three components of the velocity), by exploiting some symmetry of the system.

\subsection{Charging and shielding of a conducting planar wall in an unmagnetized plasma}

First, we study the charging and shielding of a perfectly conducting planar wall in an unmagnetized plasma consisting of electrons and singly charged ions. 
We consider a rectangular domain $L_x\times L_y=20 \times 5$ (in units of electron Debye length) in Cartesian geometry and use a uniform mesh:
\begin{eqnarray}
 &&x=L_x \xi, \nonumber \\
&&y=L_y \eta.
\label{map3}
\end{eqnarray}
The wall is located at the left boundary, $\xi=0$, and absorbs plasma. The computational domain is initially empty. At each time step we inject 
plasma particles from the right boundary $\left(\xi=1 \right)$. The injection fluxes are calculated according to a Maxwellian at rest for the 
electrons and a drifting Maxwellian (with drift velocity $v_{di}=-2.1$, normalized to the ion thermal speed $v_{thi}=\sqrt{T_i/m_i}$
with $T_i$ the ion temperature) for the ions. The boundary conditions 
for the field solver are periodic at the bottom $\left(\eta=0 \right)$ and top $\left(\eta=1 \right)$ boundaries, 
$\phi=0$ at the right boundary, and Gauss' law is applied at the wall. Particles can be absorbed by the wall (namely they are removed from the 
simulation and their charge is accumulated to the wall), can leave the system at the right boundary, while they re-enter at the bottom (top) 
if they cross the top (bottom) boundary. Other parameters of the simulation are $\omega_{pe}\Delta t=0.05$ 
(with $\omega_{pe}$ the electron plasma frequency), the temperature ratio of the injected plasma is $T_e/T_i=1$ and the mass ratio is
$m_i/m_e=1836$.

Given the plasma fluxes at the right boundary, it is possible to calculate the (one-dimensional) steady-state solution of the system. 
The theory is based on particle and energy conservation. It requires the solution of a nonlinear Poisson equation, where the plasma densities 
are obtained analytically as function of the electrostatic potential. We do not report these expressions here for brevity. A similar theory, 
which does not include the ion drift velocity, can be found in Ref. \cite{schwager}. We refer to this procedure as the analytic solution and use it to benchmark against CPIC.

\begin{figure}
\centering
\includegraphics[width=3.3in]{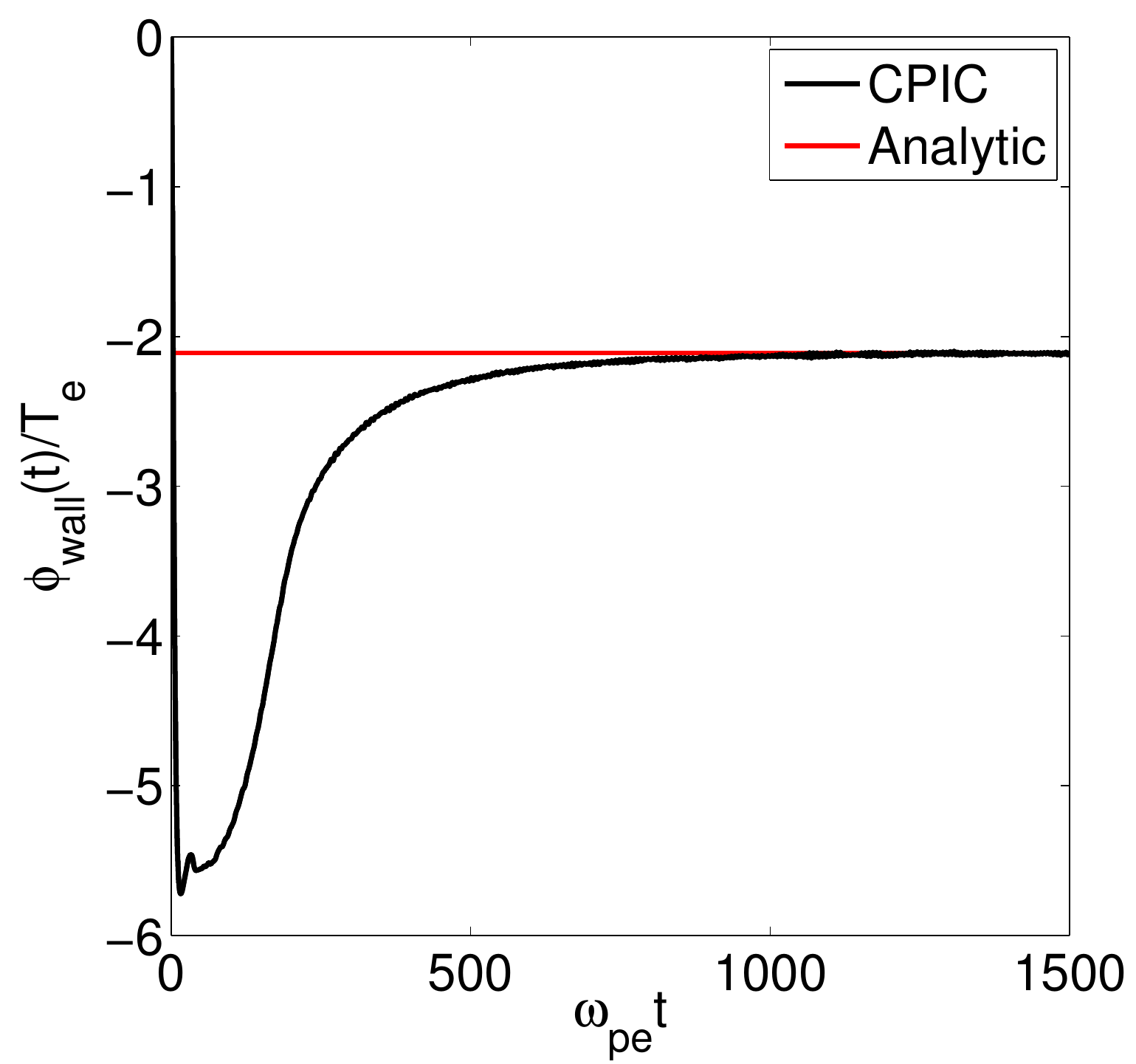}
\caption{Test A: time evolution of the wall potential.}
\label{fig1}
\end{figure}

\begin{figure}
\centering
\includegraphics[width=4in]{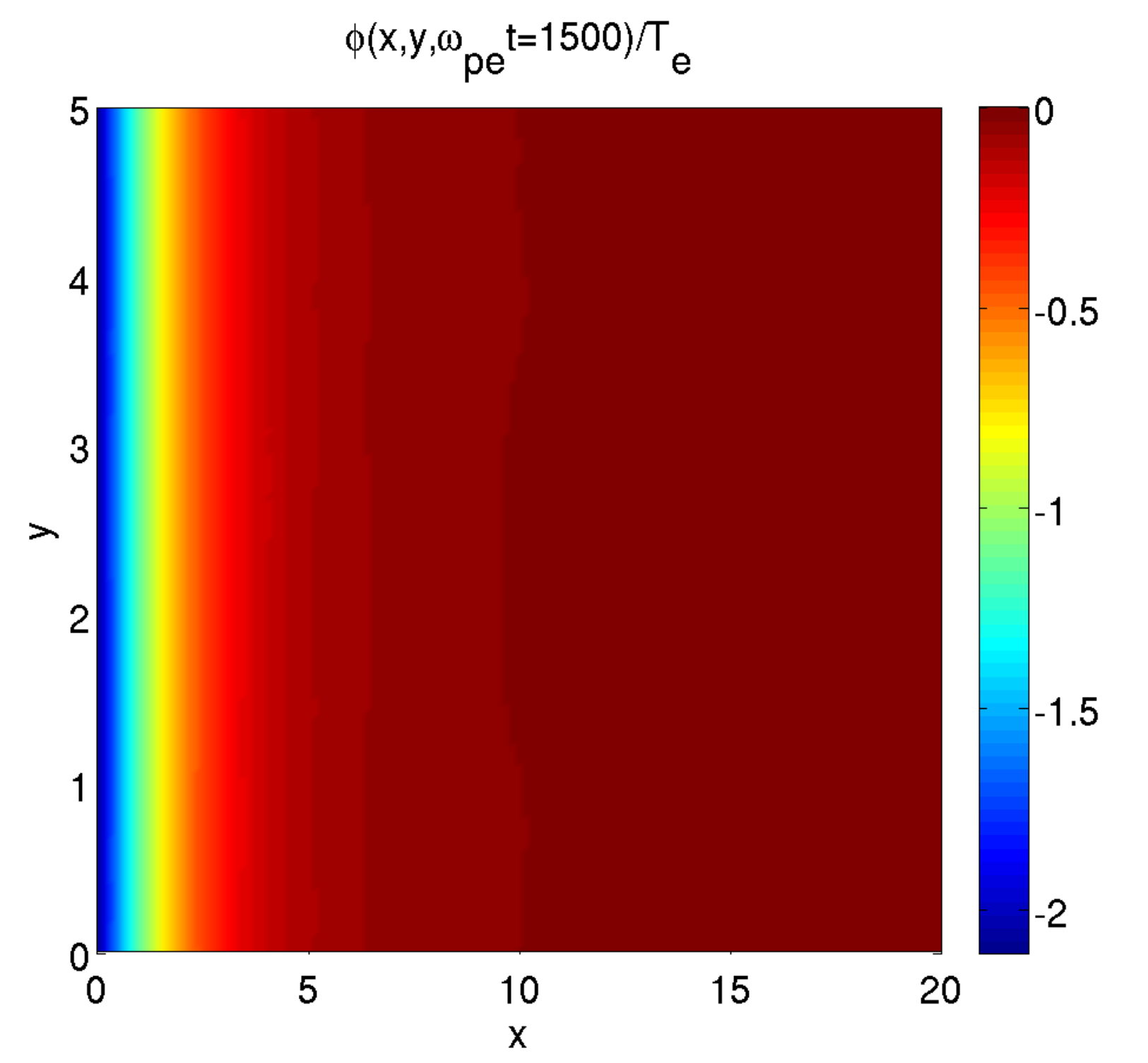}
\caption{Test A: equilibrium shielding potential at the end of the simulation.}
\label{fig2}
\end{figure}

Figure \ref{fig1} shows the time evolution of the wall potential (normalized to the electron temperature) obtained with CPIC for a $128\times128$ mesh. 
The computational domain is initially empty. As the electrons move towards the wall much faster than the ions, at the beginning there is 
a strong negative spike in the potential: $\phi_{wall}\left(\omega_{pe}t\right)/T_e\simeq-5.7$.
Slowly, the ion distribution function adjusts to its equilibrium value and a steady-state is reached. At equilibrium, the wall potential 
(averaged over the last sixth of the simulation) is $\phi_{wall}/T_e\simeq-2.11$, in good agreement with the one from the analytic theory
$\phi_{wall}^{analytic}/T_e\simeq-2.10$.
Figure \ref{fig2} shows the shielding potential at the end of the simulation. As expected, the structure of the potential is one-dimensional
(aside from some fluctuations in the region where the potential is very small).
A comparison of the equilibrium shielding potential with the analytic solution (not shown) indicates that the screening of the wall by the plasma 
is captured correctly in the simulation.

\subsection{Charging and shielding of a conducting sphere in an unmagnetized plasma}

The next natural test is to study the charging and shielding of a perfectly conducting spherical object in an unmagnetized plasma (see for instance Ref. \cite{laframboise}). 
We can exploit the curvilinear geometry of CPIC and the fact that the system has spherical symmetry. 
Thus, we run CPIC in 2D with a physical space represented in cylindrical geometry $\left(x^1=r,\,x^2=\theta,\,x^2=z\right)$,
and imposing azimuthal symmetry. The coordinate transformation between physical and logical space is then given by Eqs. (\ref{r}) and (\ref{z}). 
This corresponds to a physical domain between two spheres of radii $r_1$ and $r_2$. 

The simulation setup is as follows. The computational box is initially loaded with a Maxwellian plasma at rest. 
The boundary conditions on the left and right boundaries in logical space (inner and outer spheres, respectively) remain the same as in the previous example. 
However, we use homogeneous Neumann boundary conditions (zero normal derivative of the potential, 
$\partial \phi/\partial r=\partial \phi/\partial \xi=0$ since in this case the grid is orthogonal) for the fields at the bottom and top boundaries 
in logical space (corresponding to the $z$ axis in physical space) and 
particles are specularly reflected there. At the right boundary, we inject Maxwellian fluxes with no drift velocities. 
The other simulation parameters are the same as in Test A, except for $\omega_{pe}\Delta t=0.1$. We also set $r_1=1$ and $r_2=10$.

The analogue of the equilibrium theory briefly discussed in the previous example can be developed in spherical geometry (see for instance Ref. \cite{kennedy}). 
It is more complicated due to the fact that the conservation of angular momentum and at least two components of the particle velocity 
must be included in the calculation. For these reasons, it is common to use an approximated theory, the Orbital-Motion-Limited (OML) theory \cite{langmuir}, 
to obtain the sphere floating potential. While OML is an approximated theory, 
previous studies have shown that OML indeed provides a quite accurate result even 
when the sphere radius is comparable to the plasma Debye length \cite{delzanno1}.

\begin{figure}
\centering
\includegraphics[width=3.3in]{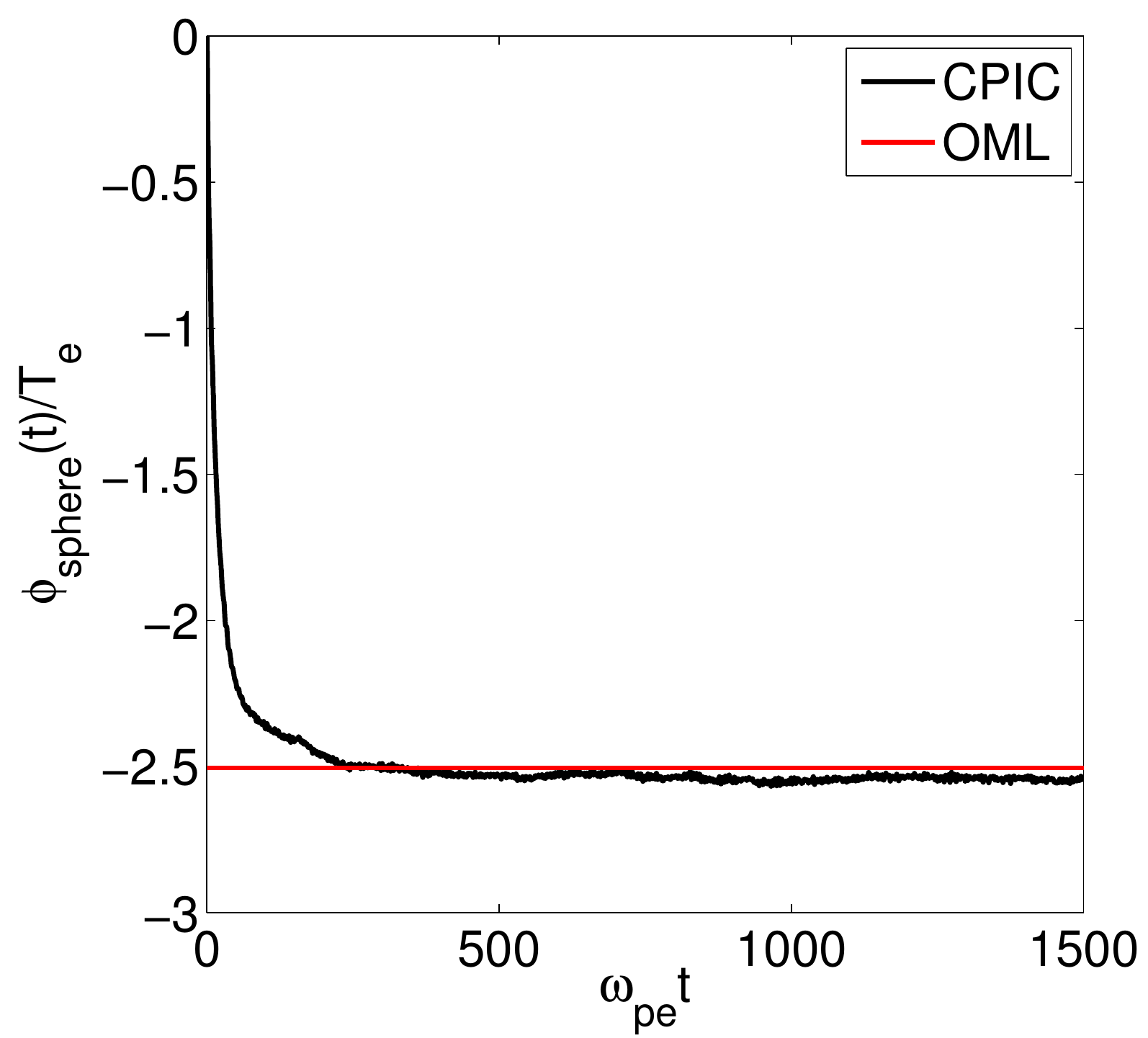}
\caption{Test B: time evolution of the wall potential.}
\label{fig3}
\end{figure}

\begin{figure}
\centering
\includegraphics[width=4in]{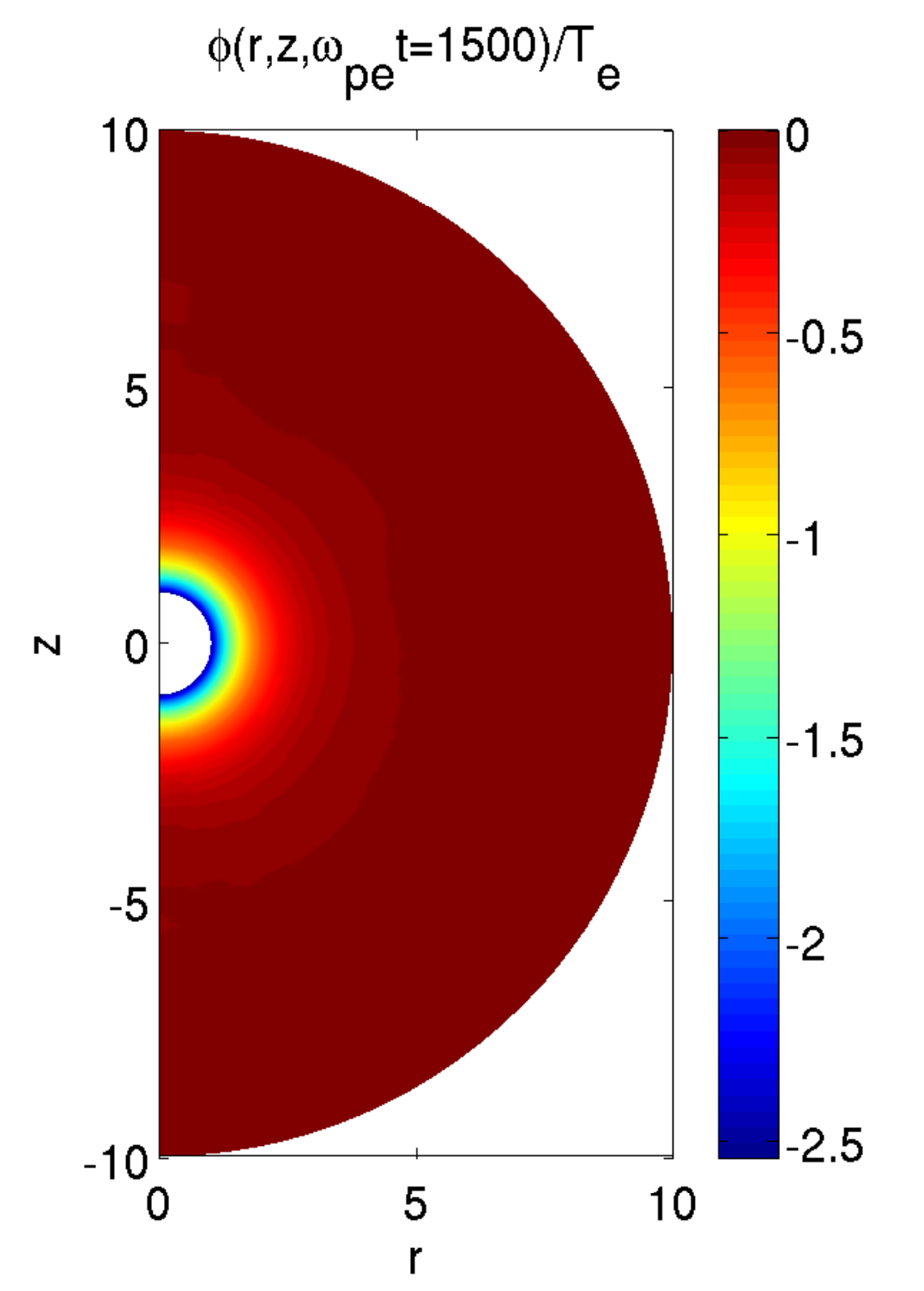}
\caption{Test B: equilibrium shielding potential at the end of the simulation.}
\label{fig4}
\end{figure}

\begin{figure}
\centering
\includegraphics[width=3.3in]{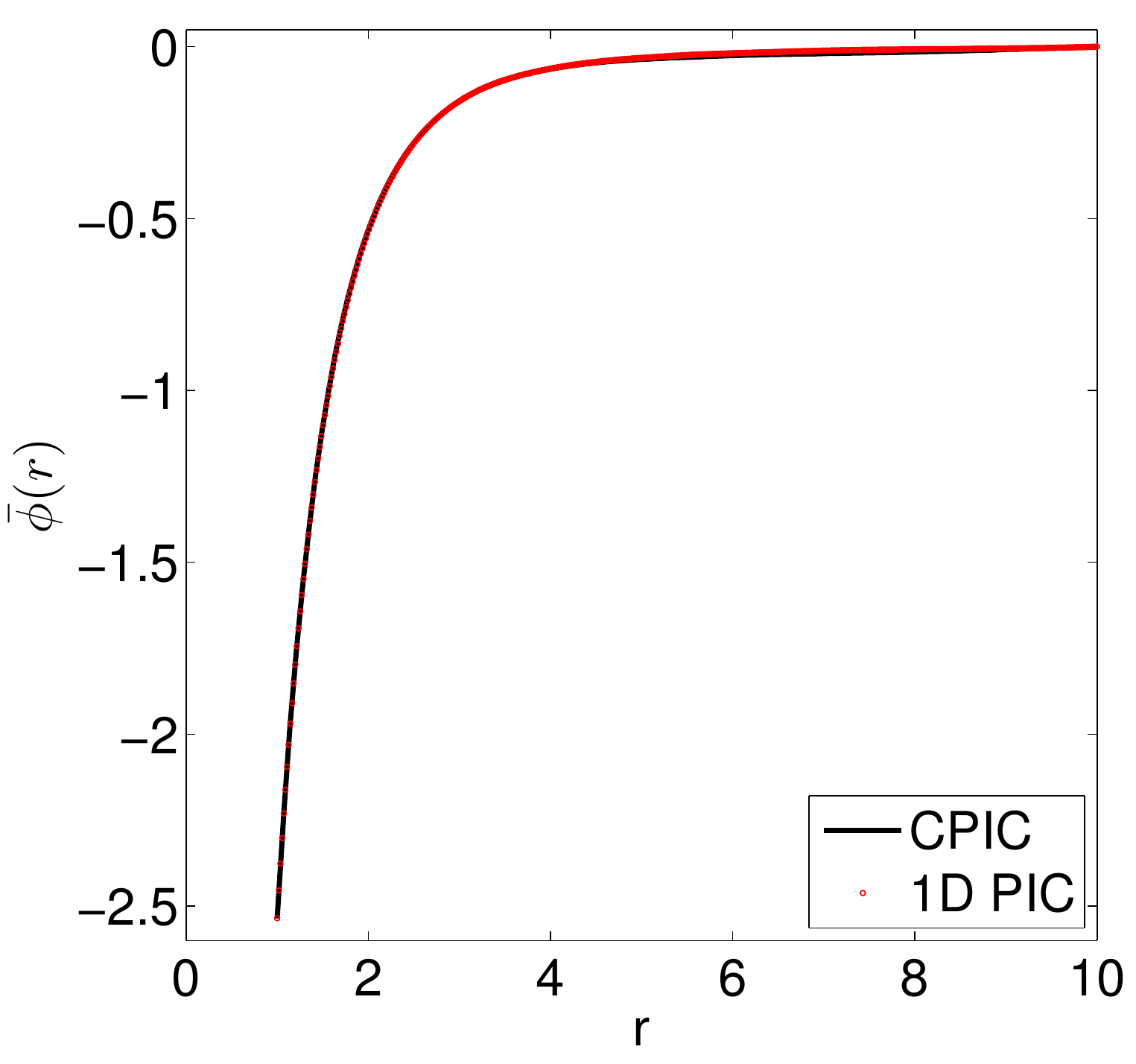}
\caption{ Test B:  equilibrium shielding potential at the end of the simulation, averaged over $\eta$.}
\label{fig5}
\end{figure}

Figure \ref{fig3} shows the time evolution of the normalized floating potential obtained for a $128\times 128$ mesh. 
Unlike the case of Fig. \ref{fig1}, the potential approaches its asymptotic equilibrium value monotonically. This is because the computational domain 
is loaded with plasma at t=0, and the ions immediately start charging the sphere. One can also see that the asymptotic equilibrium value,
$\phi_{sphere}/T_e\simeq -2.54$, matches reasonably well the equilibrium value obtained by the OML theory, $\phi_{OML}/T_e=-2.50$.
The fact that the equilibrium value is slightly lower than OML can be attributed to the development
of an absorption barrier for the ions, as discussed in Ref. \cite{daugherty}. 
The equilibrium shielding potential is shown in Fig. \ref{fig4}. The spherical symmetry is evident from the
plot. In Fig. \ref{fig5}, we compare the equilibrium shielding potential averaged over $\eta$ with that obtained with a 1D PIC code designed in 
spherical geometry. The latter was used for instance in Refs. \cite{delzanno1,delzanno2} to study charging and shielding of electron emitting spherical dust grains 
and is well benchmarked. There is a good agreement between the two, indicating that CPIC captures correctly the screening of the charged sphere by the plasma. 
Furthermore, we have performed a convergence study changing the number of grid points with the goal of checking the scalability of the solver in a 
practical situation. The results are presented in Table \ref{t4} (which also shows the average number of particles per cell at the end of the simulation). 
The number of solver iterations averaged over the entire simulation remains fairly flat.

\begin{table}
\caption{Convergence study of the BoxMG solver for Test B.}
\label{t4}
\centering
\begin{tabular}{|c|c|c|c|}
\hline
Grid & Average number& Average number & Average number\\
 & of iterations &  of electrons per cell & of ions per cell \\
\hline
\hline
$32^2$ & $8.0$ & $231$ & $330$ \\ \hline
$64^2$ & $8.0$ & $230$ & $276$ \\ \hline
$128^2$ & $8.0$ & $230$ & $248$ \\ \hline
$256^2$ & $9.0$ & $230$ & $230$ \\ \hline \hline
\end{tabular}
\end{table}

\subsection{Charging and shielding of a conducting sphere in a magnetized plasma}

Our last example involves the charging and shielding of a conducting, spherical object in a magnetized plasma.
We consider a uniform magnetic field directed along $z$.
For this case, since there is no analytic theory that relates the sphere floating potential to the plasma
parameters for arbitrary values of the magnetic field, we perform some of the studies 
of Ref. \cite{marchand} for benchmark. Specifically, we consider two cases. In the first one, the magnetic
field has magnitude $B_{z0}\simeq 1.9$ such that the electron thermal gyroradius is $\rho_e/\lambda_{De}=0.533$ and the
electron gyrofrequency is $\omega_{ce}/\omega_{pe}\simeq 1.9$. Other parameters and the simulation settings
are as in Test B. In particular, the ion thermal gyroradius is $\rho_i/\lambda_{De}\simeq 22.8$ and the ion gyrofrequency
is $\omega_{ci}/\omega_{pe}\simeq 0.001$. Thus, in this example, the electron
gyroradius is comparable to the sphere radius and we refer to this case as 'weakly magnetized'.
In the second example we increase the value of the magnetic field by a factor of $10$, $B_{z0}\simeq 19$.
Consistently, the electron thermal gyroradius is $\rho_e/\lambda_{De}=0.0533 \ll r_1/\lambda_{De}$, while the ion
thermal gyroradius is $\rho_i/\lambda_{De}\simeq 2.3>r_1/\lambda_{De}$. The electron gyrofrequency is 
$\omega_{ce}/\omega_{pe}\simeq 18.8$, while for the ions $\omega_{ci}/\omega_{pe}\simeq 0.01$.
We refer to this case as 'moderately magnetized', to distinguish from the strongly magnetized case when
$\rho_i\ll r_{1}$. We note that with these values of the magnetic field the electron gyromotion
corresponds to the shortest length scale and fastest frequency in the system.

We have performed simulations with CPIC for the systems just described using a $128 \times 128$ grid with time
step $\omega_{pe}\Delta t=0.05$. For the moderately magnetized case we note the important fact that this
choice of the time step underresolves the electron gyromotion ($\omega_{ce} \Delta t \approx 1$). A PIC code
with a standard leap-frog particle mover would be very inaccurate and possibly numerically
unstable for these parameters. As we have argued above, the reason why we can afford such
large time step is that for a uniform magnetic field CPIC uses the cyclotronic particle integrator
\cite{patacchini,delzanno_mover} and only needs to resolve
accurately the dynamics associated with the electric field.

For the weakly magnetized case, the floating potential averaged over the last sixth of the simulation is
$\phi_{sphere}/T_e\simeq -2.51$, slightly lower than the one obtained for the unmagnetized plasma. In addition,
the shielding potential remains symmetric (not shown). 
For the moderately magnetized case, the floating potential averaged
over the last sixth of the simulation is $\phi_{sphere}/T_e\simeq -2.43$, further lowered relative to the
weakly magnetized case. Also, the shielding potential is not symmetric anymore (not shown).

\begin{figure}
\centering
\includegraphics[width=3in]{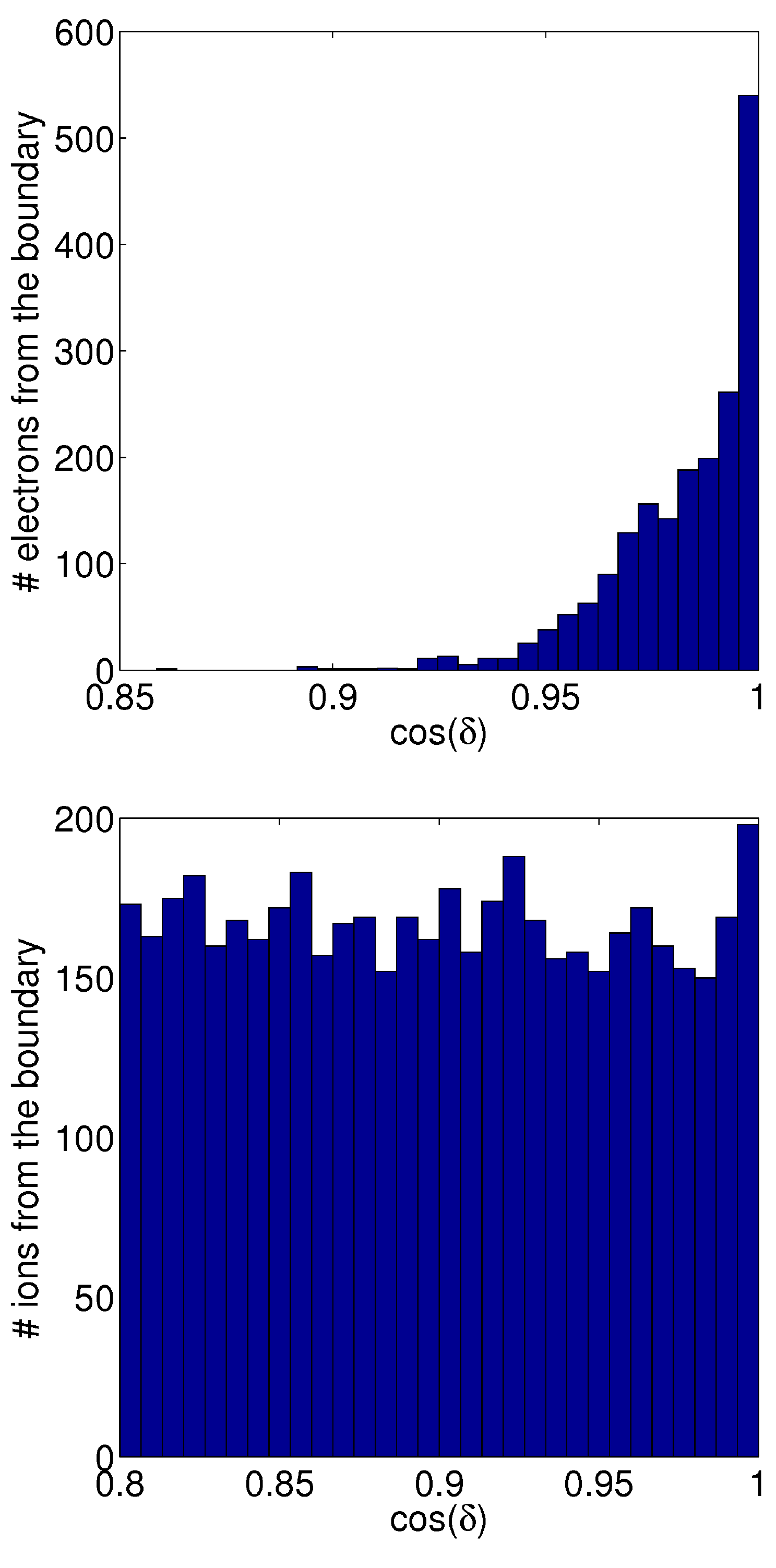}
\caption{Test C, $B_{z0}\simeq 1.9$: Outer boundary angular distribution,
$\cos(\delta)=\cos[\pi(1-\eta)]$, of the test particles that hit the sphere.
Initially $500,000$ test particles for each species were injected from the outer boundary with 
Maxwellian distribution functions. The test particles were distributed uniformly on the boundary
in the region defined by $\cos\left[\pi\left(1-\eta\right) \right]\ge0.8$.}
\label{fig6}
\end{figure}

In order to interpret the results just discussed, we have studied the trajectories of test particles
moving in the equilibrium configuration obtained from CPIC. We have injected $500,000$ particles for
each species, uniformly located on the outer boundary (in the region defined by
$\cos\left[\pi\left(1-\eta\right) \right]\ge0.8$) and with injection velocities
distributed according to a Maxwellian, and have
followed them until they either hit the sphere or leave the system.
Figure \ref{fig6} shows the number of particles that were able to hit the sphere in terms of their initial
angular distribution on the outer boundary $\cos(\delta)$, with
\begin{equation}
\delta=\pi \left(1-\eta\right).
\label{delta}
\end{equation}
The parameter $\delta$ is the colatitude in the spherical coordinate system with axis along the magnetic field:
$\cos \delta=0$ corresponds to the equatorial plane, while $\cos \delta =\pm 1$ is the sphere axis.
The plot in Fig. \ref{fig6} is obtained for the weakly magnetized case ($B_{z0}\simeq1.9$).
One can clearly see that the ions (bottom panel) can reach the sphere isotropically from the outer boundary.
On the other hand, the majority of electrons (top panel) is collected in a flux tube of characteristic radius
$r_{flux}\simeq r_1+\rho_e$ directed along the magnetic field, for which $\cos(\delta)\simeq 0.98$.
However, it is interesting to note that some electrons are collected in a larger flux tube.
For instance, $\cos(\delta)=0.9$ corresponds to a radius $r\simeq 4.4$. These are electrons that have
higher initial velocities, towards the tail of the Maxwellian distribution function, and therefore have a large gyroradius.

\begin{figure}
\centering
\includegraphics[width=3.5in]{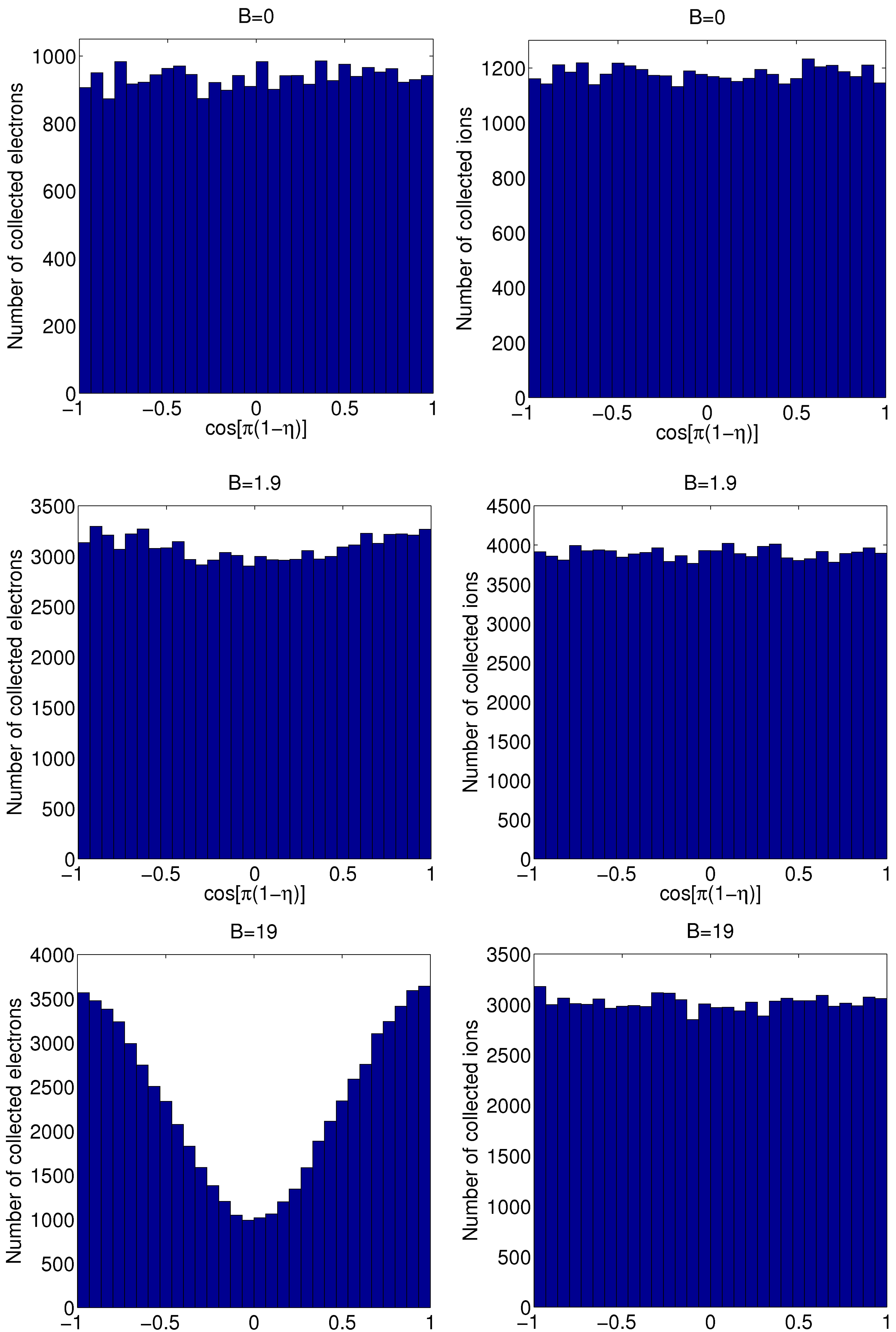}
\caption{Test C: Angular distribution,
$\cos(\delta)=\cos[\pi(1-\eta)]$, of particles collected by the sphere over the last tenth of each simulation
performed for Tests B and C.}
\label{fig7}
\end{figure}

The diagnostic just described is complemented with the distribution of particle collection
on the surface of the sphere. This is shown in Fig. \ref{fig7} with data obtained directly from
CPIC over the last tenth of each simulation in Tests B and C. 
The top panels are for the unmagnetized case, the middle panels for the weakly magnetized case
and the bottom panels for the moderately magnetized case.
As expected, when the plasma is
unmagnetized, both electrons and ions are collected isotropically on the sphere. When the plasma
is weakly magnetized the ions are still collected isotropically, since
$\rho_i\gg r_1$. The electrons start to show a small depletion around $\cos(\delta)=0$. Interestingly,
even though the electrons are collected primarily in a flux tube centered around the sphere with
characteristic radius of a few electron thermal gyroradii, they can still hit the sphere quite isotropically
since their gyroradius is comparable to the sphere radius. As a consequence, the shielding potential
remains symmetric. For the moderately magnetized case ($B_{z0}\simeq 19$), $\rho_e \ll r_1$ and the electron
motion is severely constrained by the magnetic field. Therefore, the electrons do not hit the sphere
isotropically, and there is a large depletion in the number of electrons around $\cos(\delta)\approx 0$.
The ions, on the other hand, still hit the sphere isotropically since $\rho_i>r_1$.

As a final comment, we note that these results are consistent with the 
findings of Ref. \cite{marchand}. For weakly to moderately
magnetized plasmas, the electron collection is restricted by the magnetic field and the resulting electron
current is lower relative to the unmagnetized case. Consequently, the floating potential on the sphere
becomes less negative. Quantitatively, Ref. \cite{marchand} gives $\phi_{sphere}/T_e=-2.41$ ($B_{z0}=0$),
$\phi_{sphere}/T_e=-2.37$ ($B_{z0}=1.9$), and $\phi_{sphere}/T_e=-2.21$ ($B_{z0}=19$), which is in
reasonable agreement (within $\sim 10\%$) with our results.  

\section{Conclusion}
We have described CPIC: a fully-kinetic, electrostatic, body-fitted PIC code in curvilinear geometry for structured grids. 
CPIC was designed with the goal of flexibility (namely the ability to handle different geometries, complex computational domains, and the interaction of plasmas with complex objects with characteristic size smaller than the characteristic 
plasma length scales), robustness and performance. It introduces a coordinate transformation from the physical space to the logical 
space, where the grid is uniform and Cartesian. In CPIC, most operations are performed in logical space. Its main features are 
(1) the use of structured grids, (2) a 
scalable and robust field solver based on the Black Box multigrid algorithm, and (3) an hybrid particle mover, where the particles' 
position is updated in logical space while the particles' velocity is updated in physical space. In our tests, the hybrid mover 
has proven more efficient and robust than the conventional physical space mover. We have presented some successful benchmark tests 
involving the interaction of a plasma with material boundaries, in Cartesian and spherical geometries, with
and without magnetic field.

\section*{Acknowledgment}
The authors wish to thank Luis Chacon, Chris Fichtl, John M. Finn, Rao Garimella, Richard Marchand, and Xianzhu Tang. 
This research is supported by the LDRD program of the Los Alamos National Laboratory and was performed under 
the auspices of the NNSA of the U.S. DOE by LANL, operated by LANS LLC under Contract No. DE-AC52-06NA25396.

{\bibliographystyle{ieeetr}
\bibliography{articolo_final_nophoto}}

\begin{thebibliography}{10}

\bibitem{birdsall}
C.~K. Birdsall and A.~B. Langdon, {\em Plasma Physics Via Computer Simulation}.
\newblock Taylor \& Francis, 2004.

\bibitem{hockney}
R.~W. Hockney and J.~W. Eastwood, {\em Computer Simulation Using Particles}.
\newblock Taylor \& Francis, 1988.

\bibitem{brackbill&forslund}
J.~Brackbill and D.~Forslund, ``An implicit method for electromagnetic plasma
  simulation in two dimensions,'' {\em Journal of Computational Physics},
  vol.~46, no.~2, pp.~271 -- 308, 1982.

\bibitem{chacon}
G.~Chen, L.~Chacon, and D.~C. Barnes, ``An energy- and charge-conserving,
  implicit, electrostatic particle-in-cell algorithm,'' {\em Journal of
  Computational Physics}, vol.~230, no.~18, pp.~7018--7036, 2011.

\bibitem{markidis}
S.~Markidis and G.~Lapenta, ``The energy conserving particle-in-cell method,''
  {\em Journal of Computational Physics}, vol.~230, no.~18, pp.~7037--7052,
  2011.

\bibitem{roussel98}
J.-F. Roussel, ``Spacecraft plasma environment and contamination simulation
  code: Description and first tests,'' {\em Journal of Spacecraft and Rockets},
  vol.~35, pp.~205--211, Mar. 1998.

\bibitem{hutchinson2}
L.~Patacchini and I.~H. Hutchinson, ``Spherical conducting probes in finite
  debye length plasmas and e$\times$ b fields,'' {\em Plasma Physics and
  Controlled Fusion}, vol.~53, no.~2, p.~025005, 2011.

\bibitem{vay}
J.-L. Vay, P.~Colella, J.~Kwan, P.~McCorquodale, D.~Serafini, A.~Friedman,
  D.~Grote, G.~Westenskow, J.-C. Adam, A.~Heron, and I.~Haber, ``Application of
  adaptive mesh refinement to particle-in-cell simulations of plasmas and
  beams,'' {\em Physics of Plasmas}, vol.~11, no.~5, pp.~2928--34, 2004.

\bibitem{westermann1}
T.~Westermann, ``Numerical modelling of the stationary maxwell-lorentz system
  in technical devices,'' {\em International Journal of Numerical Modelling:
  Electronic Networks, Devices and Fields}, vol.~7, no.~1, pp.~43--67, 1994.

\bibitem{westermann2}
T.~Westermann, ``Particle-in-cell simulations with moving boundaries-adaptive
  mesh generation,'' {\em Journal of Computational Physics}, vol.~114, no.~2,
  pp.~161--75, 1994.

\bibitem{munz}
C.-D. Munz, R.~Schneider, E.~Sonnendrucker, E.~Stein, U.~Voss, and
  T.~Westermann, ``A finite-volume particle-in-cell method for the numerical
  treatment of maxwell-lorentz equations on boundary-fitted meshes,'' {\em
  International Journal for Numerical Methods in Engineering}, vol.~44, no.~4,
  pp.~461--87, 1999.

\bibitem{westermann3}
T.~Westermann, ``Localization schemes in 2d boundary-fitted grids,'' {\em
  Journal of Computational Physics}, vol.~101, no.~2, pp.~307--313, 1992.

\bibitem{eastwood}
J.~W. Eastwood, W.~Arter, N.~J. Brealey, and R.~W. Hockney, ``Body-fitted
  electromagnetic pic software for use on parallel computers,'' {\em Computer
  Physics Communications}, vol.~87, no.~1-2, pp.~155--178, 1995.

\bibitem{fichtl}
C.~A. Fichtl, J.~M. Finn, and K.~L. Cartwright, ``An arbitrary
  curvilinear-coordinate method for particle-in-cell modeling,'' {\em
  Computational Science and Discovery}, vol.~5, no.~1, p.~014011 (25 pp.),
  2012.

\bibitem{wang}
J.~Wang, D.~Kondrashov, P.~Liewer, and S.~Karmesin, ``Three-dimensional
  deformable-grid electromagnetic particle-in-cell for parallel computers,''
  {\em Journal of Plasma Physics}, vol.~61, no.~3, pp.~367--389, 1999.

\bibitem{mandell}
M.~J. Mandell, V.~A. Davis, D.~L. Cooke, A.~T. Wheelock, and C.~J. Roth,
  ``Nascap-2k spacecraft charging code overview,'' {\em IEEE Transactions on
  Plasma Science}, vol.~34, no.~5, pp.~2084--2093, 2006.

\bibitem{roussel}
J.-F. Roussel, F.~Rogier, G.~Dufour, J.-C. Mateo-Velez, J.~Forest, A.~Hilgers,
  D.~Rodgers, L.~Girard, and D.~Payan, ``Spis open-source code: methods,
  capabilities, achievements, and prospects,'' {\em IEEE Transactions on Plasma
  Science}, vol.~36, no.~5, pp.~2360--8, 2008.

\bibitem{marchand}
R.~Marchand, ``Ptetra, a tool to simulate low orbit satellite-plasma
  interaction,'' {\em IEEE Transactions on Plasma Science}, vol.~40, no.~2 PART
  1, pp.~217--229, 2012.

\bibitem{muranaka}
T.~Muranaka, S.~Hosoda, J.-H. Kim, S.~Hatta, K.~Ikeda, T.~Hamanaga, M.~Cho,
  H.~Usui, H.~Ueda, K.~Koga, and T.~Goka, ``Development of multi-utility
  spacecraft charging analysis tool (muscat),'' {\em IEEE Transactions on
  Plasma Science}, vol.~36, no.~5, pp.~2336--49, 2008.

\bibitem{maclachlan}
S.~P. MacLachlan, J.~M. Tang, and C.~Vuik, ``Fast and robust solvers for
  pressure-correction in bubbly flow problems,'' {\em J. Comput. Phys.},
  vol.~227, pp.~9742--9761, Dec. 2008.

\bibitem{dendy}
J.~E. Dendy, Jr., ``Black box multigrid,'' {\em J. Comput. Phys.}, vol.~48,
  no.~3, pp.~366--386, 1982.

\bibitem{dendy2}
J.~E. Dendy, Jr., ``Black box multigrid for periodic and singular problems,''
  {\em Appl. Math. Comput.}, vol.~25, pp.~1--10, Jan. 1988.

\bibitem{moulton}
J.~E. Dendy, Jr. and J.~D. Moulton, ``Black box multigrid with coarsening by a
  factor of three,'' {\em Numerical Linear Algebra with Applications}, vol.~17,
  no.~2-3, pp.~577--598, 2010.

\bibitem{liseikin}
V.~D. Liseikin, {\em Grid Generation Methods}.
\newblock Berlin, New York: Springer, 1999.

\bibitem{winslow-unpublished-var-diff}
A.~Winslow, ``Adaptive mesh zoning by the equipotential method.'' UCID-19062,
  Lawrence Livermore National Laboratory, 1981.

\bibitem{delzanno08}
G.~L. Delzanno, L.~Chacon, J.~M. Finn, Y.~Chung, and G.~Lapenta, ``An optimal
  robust equidistribution method for two-dimensional grid adaptation based on
  monge-kantorovich optimization,'' {\em Journal of Computational Physics},
  vol.~227, no.~23, pp.~9841--64, 2008.

\bibitem{hypre}
{\em {HYPRE} {H}igh {P}erformance {P}reconditioners {U}ser's {M}anual, Version
  2.9.0b}.
\newblock Center for Applied Scientific Computing: Lawrence Livermore National
  Laboratory, 2012.

\bibitem{lapenta}
G.~Lapenta, ``Democritus: An adaptive particle in cell (pic) code for
  object-plasma interactions,'' {\em Journal of Computational Physics},
  vol.~230, no.~12, pp.~4679--95, 2011.

\bibitem{Briggs}
W.~L. Briggs, {\em A Multigrid Tutorial}.
\newblock SIAM, Philadelphia, 1987.

\bibitem{roache}
P.~J. Roache, ``Code verification by the method of manufactured solutions,''
  {\em Journal of Fluids Engineering, Transactions of the ASME}, vol.~124,
  no.~1, pp.~4--10, 2002.

\bibitem{boris}
J.~P. Boris, ``Relativistic {P}lasma {S}imulation - {O}ptimization of a
  {H}ybrid {C}ode,'' in {\em Proceedings of the Fourth Conference on Numerical
  Simulation of Plasmas}, (Naval Research Laboratory, Washington D.C.),
  pp.~3--67, 1970.

\bibitem{patacchini}
L.~Patacchini and I.~Hutchinson, ``Explicit time-reversible orbit integration
  in {P}article {I}n {C}ell codes with static homogeneous magnetic field,''
  {\em Journal of Computational Physics}, vol.~228, no.~7, pp.~2604 -- 2615,
  2009.

\bibitem{delzanno_mover}
G.~L. Delzanno and E.~Camporeale, ``On particle movers in cylindrical geometry
  for particle-in-cell simulations,'' {\em Journal of Computational Physics},
  vol.~253, no.~0, pp.~259 -- 277, 2013.

\bibitem{schwager}
L.~A. Schwager and C.~K. Birdsall, ``Collector and source sheaths of a finite
  ion temperature plasma,'' {\em Physics of Fluids B}, vol.~2, no.~5,
  pp.~1057--1068, 1990.

\bibitem{laframboise}
J.~G. Laframboise, ``Theory of spherical and cylindrical langmuir probes in a
  collision less {M}axwellian plasma at rest,'' {\em Ph.D. dissertation,
  University of Toronto}, 1966.

\bibitem{kennedy}
R.~Kennedy and J.~Allen, ``The floating potential of spherical probes and dust
  grains. ii: orbital motion theory,'' {\em Journal of Plasma Physics},
  vol.~69, no.~6, pp.~485--506, 2003.

\bibitem{langmuir}
H.~Mott-Smith and I.~Langmuir, ``Theory of collectors in gaseous discharges,''
  {\em Physical Review}, vol.~28, pp.~727--763, 1926.

\bibitem{delzanno1}
G.~L. Delzanno, G.~Lapenta, and M.~Rosenberg, ``Attractive potential around a
  thermionically emitting microparticle,'' {\em Phys. Rev. Lett.}, vol.~92,
  p.~035002, Jan 2004.

\bibitem{daugherty}
J.~E. Daugherty, R.~K. Porteous, M.~D. Kilgore, and D.~B. Graves, ``Sheath
  structure around particles in low-pressure discharges,'' {\em Journal of
  Applied Physics}, vol.~72, no.~9, pp.~3934--42, 1992.

\bibitem{delzanno2}
G.~L. Delzanno, A.~Bruno, G.~Sorasio, and G.~Lapenta, ``Exact orbital motion
  theory of the shielding potential around an emitting, spherical body,'' {\em
  Physics of Plasmas}, vol.~12, no.~6, p.~62102, 2005.

\end{thebibliography}

\end{document}